# Article Processing Charges based publications: to which extent the price explains scientific impact?

Abdelghani Maddi[1], David Sapinho[2]

[1] *abdelghani.maddi@hceres.fr*
Observatoire des Sciences et Techniques, Hcéres, 2 Rue Albert Einstein, Paris, 75013 France.
Sorbonne Paris Nord, CEPN, UMR-CNRS 723, Villetaneuse, 93420 France.

[2] *david.sapinho@hceres.fr*
Observatoire des Sciences et Techniques, Hcéres, 2 Rue Albert Einstein, Paris, 75013 France

**Abstract**
The present study aims to analyze relationship between Citations Normalized Score (NCS) of scientific publications and Article Processing Charges (APCs) amounts of Gold Open access publications. To do so, we use APCs information provided by OpenAPC database and citations scores of publications in the Web of Science database (WoS). Database covers the period from 2006 to 2019 with 83,752 articles published in 4751 journals belonging to 267 distinct publishers. Results show that contrary to this belief, paying dearly does not necessarily increase the impact of publications. First, large publishers with high impact are not the most expensive. Second, publishers with the highest APCs are not necessarily the best in terms of impact. Correlation between APCs and impact is moderate. Otherwise, in the econometric analysis we have shown that publication quality is strongly determined by journal quality in which it is published. International collaboration also plays an important role in citations score.

**Introduction**

Since the start of the 21st century, scientific community has witnessed an unprecedented rise in the Open Access (OA) movement (Björk, 2004; Chi Chang, 2006; Sotudeh and Estakhr, 2018). OA is seen as a good means of ensuring better dissemination of knowledge and more equity between actors, faced with the issue of paying subscription fees (Prosser, 2003; Tananbaum, 2003; Solomon and Björk, 2012; Cary and Rockwell, 2020).

However, OA does not necessarily mean "free", and raises the question of business model underlying scientific publication. For institutions, it may even generate new costs: in addition to the subscription fees, they are increasingly led to pay costs of OA publication (Maddi, 2020). This concerns a part of Gold OA publications which are based on the "author-pays" business model (Rizor and Holley, 2014; Sotudeh, Ghasempour and Yaghtin, 2015). Thus, authors pay the "Article Processing Charges" (APCs), usually via their institution, to allow open access to the publication (Asai, 2019; Khoo, 2019; Bruns, Rimmert and Taubert, 2020; Copiello, 2020). The APCs have increased significantly over time. This rise has been estimated at three times faster than it would be if indexed to inflation (Khoo 2019). The trend appears to be stronger in more frequently cited journals, as highlighted for Biomed Central journals (Asai 2020), medical and specific OA journals (Asai 2019). These findings would also suggest that large subscription journal publishers tend to set higher APCs. Nevertheless, there is no evidence to date that the introduction of APCs for a given journal reduced its publications volume (Khoo, 2019). In other words, once able to pay an APC, authors give little emphasis to their amount. The APC-based publishing model is being more and more integrated by academic institutions that aspire to switch to an economic model excluding subscription fees. Thus, APCs are now a considerable burden on "total cost of publication" for institutions, reaching 10% in 2013 (Pinfield, Salter and Bath, 2016).

---

[1] Corresponding author : Abdelghani Maddi, Observatoire des Sciences et Techniques, Hcéres, 2 Rue Albert Einstein, Paris, 75013, France, T. 33 (0)1 55 55 61 48.

While many studies have described relationships between OA publications and citation level (Gumpenberger, Ovalle-Perandones and Gorraiz, 2013; Zhang and Watson, 2017; Piwowar *et al.*, 2018), only few have focused on the amount of APCs, with heterogeneous findings. On one hand, APCs based journals have been regarded, in general, more cited than other OA journals (Björk and Solomon, 2012), on the other hand, it was concluded that both categories have, on average, similar performances with some disciplinary differences (Ghane, Niazmand and Sabet Sarvestani, 2020). Another study analyzed the relationship between APCs and scientific impact of publications using respectively DOAJ and Scopus data (Björk and Solomon, 2015). On a set of 61,081 publications and 595 journals, authors showed that there is a moderate correlation (0.4) between the two indicators at the journal level (APCs and impact). Correlation is greater (0.6) when data is weighted by the volume of articles for each journal (article level), suggesting that publishers take quality into account when pricing their journals. Likewise, authors are also sensitive to journal quality in their submission choices.

The present study aims to analyze relationship between citations normalized score of scientific publications and APCs amounts. To do so, we use APCs information provided by OpenAPC database (https://treemaps.intact-project.org/) and citations scores in the Web of Science database (WoS). Database covers the period from 2006 to 2019 with 109,141 publications (March 1, 2020). Among these publications, 83,752 match with the WoS database using DOI. Our database contains 4,751 journals with a contrasted number of publications per journal. These journals belong to 267 distinct publishers.

To the extent that large, high impact publishers/journals may request high APCs, it is expected that quality will be strongly correlated with the APCs. The latter would therefore explain the publications visibility.

**Data**

*APCs data*

APCs data has been extracted from "OpenAPC" database. It is an initiative that involves 231 institutions worldwide (5 from North America, 255 from Europe and 1 from East Asia) that publish data sets on fees paid for OA journal articles under an open database license. At the beginning of March 2020, the database contains 109,141 publications and 6,941 journals.

Each publication in this database is only assigned to the institution that declared, skipping, therefore, other collaborating institutions. For more details about Open APC database see: https://treemaps.intact-project.org/page/about.html

*OST data*

The data about citations scores and disciplinary assignation of publications has been extracted from the french Observatoire des Sciences et Techniques' (OST) in-house database. It includes five indexes of WoS available from Clarivate Analytics (SCIE, SSCI, AHCI, CPCI-SSH and CPCI-S) and corresponds to WoS content indexed through the end of March 2019. See https://clarivate.com/webofsciencegroup/solutions/webofscience-platform/.

*Final database*

The database used for analysis includes several information about publications from the two data sources:

- From the APC data : institution that declares the publication, APCs amounts, journal in which they are published, country of the journal, publisher of journal and a flag indicating whether journal is hybrid. By matching the "OpenAPC" database to that of OST.

- From the WoS-OST in-house data : we estimate the impact of publications by calculating the following indicators :
    - Normalized Citations Score (NCS) at article level: the NCS of a given article was calculated by dividing the number of citations received by the average number of citations in the same disciplines and the same year (Waltman *et al.*, 2011).
    - Mean Normalized Citation Score (MNCS) at journal and publisher level: the MNCS for a given publisher was calculated as the weighted average of the NCS scores, based on all the articles of journals that it publishes. In the case of the OpenAPC database, the selection was restricted to OA articles only.
    - Mean Normalized Impact of Journals (MNIJ) at journal and publisher level: For a given journal, the MNIJ is calculated as the average number of citations per article in a given discipline for a given year, normalized by the number of citations per article in the same discipline and the same year at the global level. The overall MNIJ of a journal (all disciplines combined) is obtained by calculating a weighted average of the MNIJs by discipline.
- Finally, international collaboration was measured by number of countries involved in publication.

The two datasets were merged on the basis of DOI, resulting in a sample of 83,752 publications.

**Method**

Spearman correlation test and regression analysis are performed to highlight the relationships between amount of APC and citations.

*Dependent variable and model choice*

The dependent variable is the logarithm of Normalized Citations Score (labelled Log (NCS)) received by each publication during the period 2006-2019. To retain the zeros, we have added 1 to the NCS before making the logarithmic transformation. Log (NCS) is a continuous variable with a lower boundary at zero and an upper boundary at infinity. Thus, a left censored Tobit regression model is used to account for the disproportionate number of observations with zero values, because a significant proportion of the observations in our sample are zeros. Tobit regressions avoid inconsistent estimates from OLS regression.

*Explanatory variable*

In this study, we seek to analyze to what extent the amount of APCs have an incidence on the number of citations received by OA scientific publications. Our explanatory variable is therefore the amount of APCs by publication.

*Control variables*

Journal impact and number of countries per publication are added as control variables. This choice was driven by the literature that shows that citations depend on journal quality and international collaboration (Maddi, Larivière and Gingras, 2019; Maddi and Gingras, 2021). The hybrid status of a journal was also observed, using a dummy variable.

**Results**

In this section we present the main results. First, we characterize the APCs data, namely: evolution of the average amount paid by institutions, characteristics of the top 20 producing publishers and then those of the most expensive one. Secondly, we present the correlation tests results between the amount of APCs on the one hand and the MNCS and MNIJ indicators on the other hand. Finally, we present results from regression.

*Overview of APCs data*

Figure 1 shows the evolution of APCs average by publication in all "OpenAPC" database, and using a constant data set of publications from journals of 2006-09 period (79 journals). As we can see, the average amount has doubled between 2006 and 2019, going from 1,000 euros to almost 2,300 euros per publication. This is explained in particular by the new journals that have been indexed into OpenAPC database after 2009, which significantly increases the APCs average. Overall, APCs have increased significantly even for old journals (2006-09) from 1,000 to 1,800 euros on average.

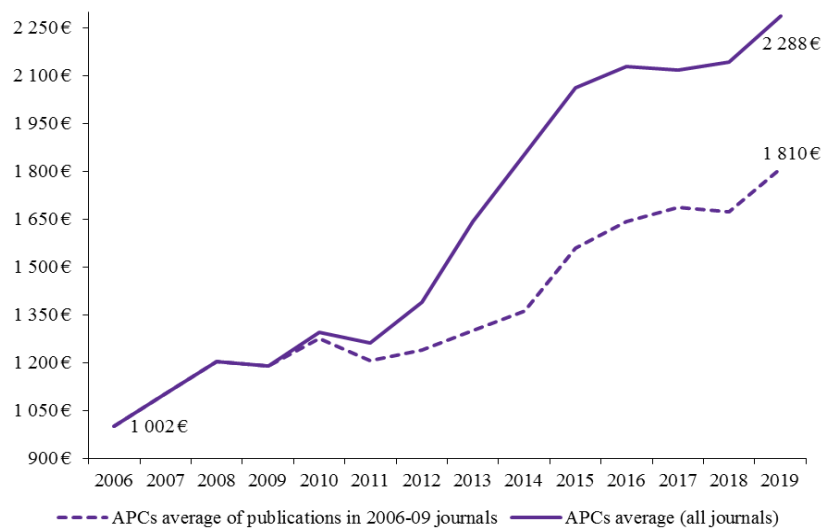

**Figure 1: APCs Average per publication, total and with constant journal set**

With all reservations that we can make on the Open APC database, we can hypothesize that this increase is notably due to a rise in demand for OA publication. As is well known in economics, increase in demand systematically leads to an increase in prices. Publishers are therefore taking advantage of this enthusiasm for OA to increase their prices.

Figure 2 describes the distribution of publications in the OpenAPC database, in relation to the amount of APCs, by discipline.

First, it highlights contrasting gold open access publication practices. This finding is similar to that observed from the WoS data base for OA publications in previous studies (Maddi, 2020; Maddi, Lardreau and Sapinho, 2021). However, the distribution is quite different than that of all the publications where the weight of engineering or chemistry, for example, is higher (OST, 2019). Thus, nearly 50% of the publications in the OpenAPC database are in Fundamental biology and Medical research, with respectively 23,466 and 19,669 publications. These two disciplines account for only 30% of publications in the WoS database (and 42% of Gold open access). The least present disciplines are mathematics, computer science and humanities. This result is largely explained by the overall size of these disciplines (see OST, 2019).

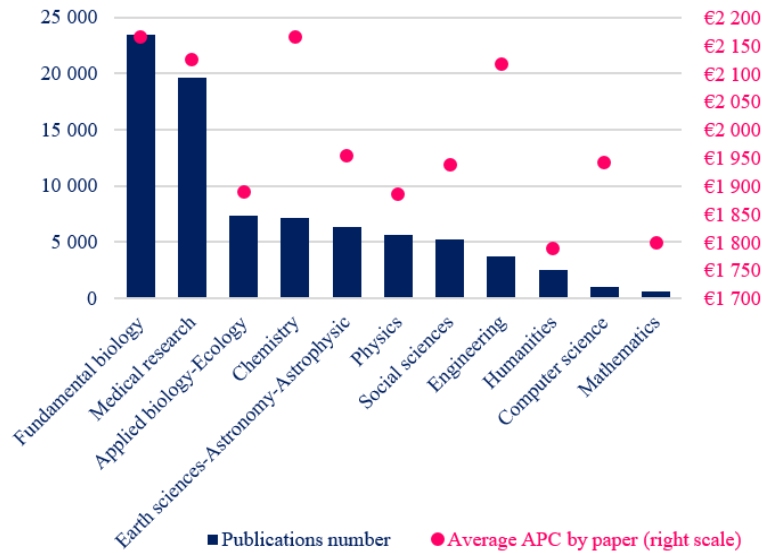

**Figure 2: publications number and average APC by discipline**

The average amount of APCs by discipline varies from € 1,800 (Mathematics and Humanities) to € 2,150 (Fundamental biology and Chemistry). The difference is therefore not significantly high depending on the discipline.

Figure 3 shows the distribution of publications in the OpenAPC database (top 20 countries) as well as the corresponding average APC amount per country.

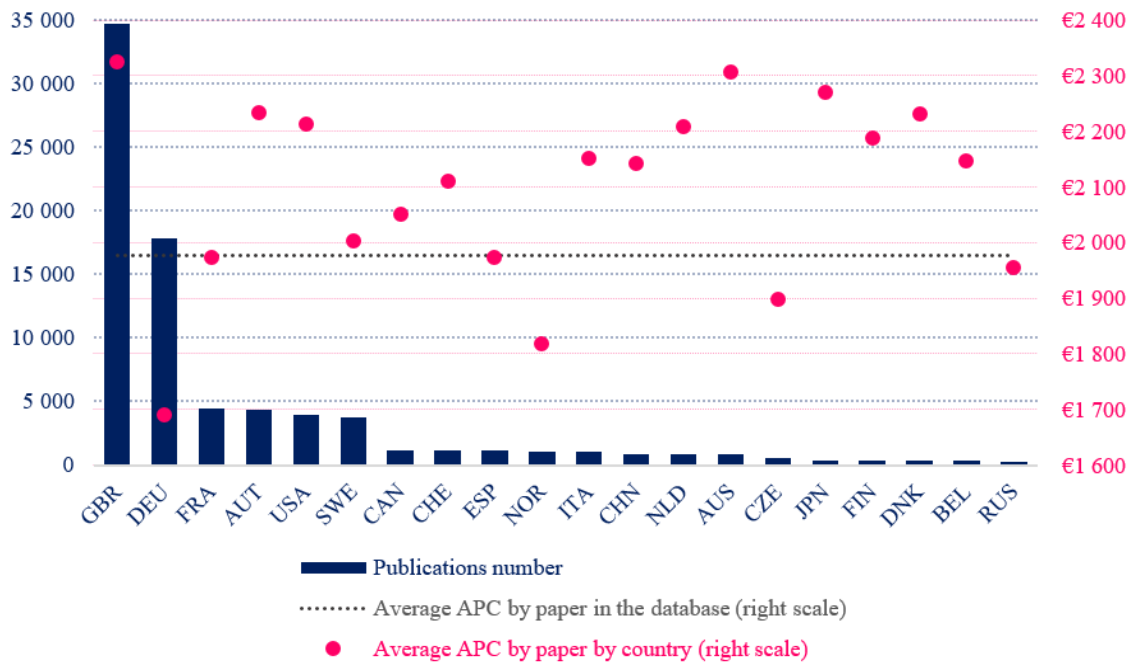

**Figure 3: publications number and average APC by country (Top 20 countries in OpenAPC database)**

The UK accounts for 40% of the OpenAPC database publications with almost 35,000 articles (fractional count), followed by far by Germany with 17,779 articles. France comes in third position, with 4,450 publications, followed closely by Austria, the United States and Sweden. This distribution shows that the OpenAPC database is biased in favor of European countries. The scope of our results is therefore limited in particular to European countries. When it comes to the average amount of APC per country, the UK also comes first with an average APC of

2325 euros. In contrast, Germany, the second producer country in the OpenAPC database, has the lowest average APC per paper (1690 euros), followed by Norway with average APCs of 1817 euoros.

**Table 1: top 20 producing publishers: publications numbers, APCs average, MNIJ and MNCS**

| Publisher | # publications | APCs average | MNIJ | MNCS |
|---|---|---|---|---|
| Springer Nature | 14103 | 1 992 € | 1,75 | 1,29 |
| Elsevier BV | 12534 | 2 855 € | 1,99 | 1,99 |
| Public Library of Science (PLoS) | 9027 | 1 448 € | 1,46 | 1,04 |
| Wiley-Blackwell | 6959 | 2 351 € | 1,78 | 1,63 |
| Frontiers Media SA | 5725 | 1 686 € | 1,25 | 0,95 |
| MDPI AG | 3438 | 1 212 € | 1,16 | 0,85 |
| Springer Science + Business Media | 3313 | 1 536 € | 1,45 | 1,23 |
| Oxford University Press (OUP) | 3022 | 2 411 € | 2,33 | 1,97 |
| American Chemical Society (ACS) | 2299 | 2 627 € | 3,48 | 1,81 |
| IOP Publishing | 2127 | 1 569 € | 1,55 | 1,37 |
| Copernicus GmbH | 1994 | 1 492 € | 1,96 | 1,38 |
| Informa UK Limited | 1911 | 1 390 € | 0,87 | 1,47 |
| BMJ | 1604 | 2 089 € | 0,97 | 1,76 |
| Royal Society of Chemistry (RSC) | 1131 | 1 629 € | 2,49 | 1,22 |
| Optical Society of America (OSA) | 905 | 1 891 € | 1,72 | 1,91 |
| SAGE Publications | 829 | 929 € | 0,85 | 1,55 |
| Institute of Electrical & Electronics Engineers (IEEE) | 803 | 1 505 € | 1,82 | 3,12 |
| The Royal Society | 774 | 1 926 € | 1,80 | 1,48 |
| Hindawi Publishing Corporation | 739 | 1 370 € | 0,67 | 0,53 |
| Ovid Technologies (Wolters Kluwer Health) | 715 | 3 215 € | 1,41 | 1,82 |

Table 1 shows that more than half of publications are concentrated on the top 5 publishers. With a few exceptions, APCs are lower for large publishers than for the overall average. The MNIJ is higher than the world average for almost all publishers. Likewise for the MNCS. Furthermore, Table 1 shows that for some publishers, impact of publications (MNCS) is much higher than that of journals (MNIJ). This is particularly the case for the publishers "Informa UK Limited" and "BMJ". Explanation can be found in the fact that the majority of journals for these publishers (respectively, 87 and 93%) are either closed or hybrid. As demonstrated in the literature, OA publications are more cited than non-OA ones. Consequently, MNCS of publications indexed in OpenAPC database would be systematically higher than the average impact of journals in which they are published.

Table 2 shows that among the largest publishers, only Elsevier BV is listed in the top 20 most expensive ones (20th position), that are mostly American. We can also note that both impact of publications and impact of journals are high. Some exceptions can be made, especially for "MyJove Corporation" publisher with an average amount of 3081 euros for a very low impact. Similarly, "The American Association of Immunologists" charges for expensive APCs, while the average impact of journals and publications is at the level of the world average.

**Table 2: top 20 most expensive publishers: publications numbers, APCs average, MNIJ and MNCS**

| Publisher | # publications | APCs average | MNIJ | MNCS |
|---|---|---|---|---|
| American Society for Nutrition | 47 | 4 761 € | 2,34 | 2,32 |
| American Medical Association (AMA) | 28 | 4 624 € | 2,70 | 5,73 |
| American Society of Clinical Oncology (ASCO) | 23 | 4 588 € | 1,35 | 2,63 |
| Rockefeller University Press | 66 | 4 466 € | 3,64 | 2,32 |
| American Psychological Association (APA) | 132 | 3 754 € | 1,45 | 1,86 |
| American Society for Clinical Investigation | 96 | 3 656 € | 3,49 | 2,80 |
| Royal College of Psychiatrists | 64 | 3 630 € | 1,45 | 1,98 |
| EMBO | 162 | 3 410 € | 3,13 | 2,29 |
| European Respiratory Society (ERS) | 29 | 3 293 € | 1,44 | 1,80 |
| Ovid Technologies (Wolters Kluwer Health) | 715 | 3 215 € | 1,41 | 1,82 |
| American Association for Cancer Research (AACR) | 73 | 3 175 € | 2,05 | 1,45 |
| Nature Publishing Group | 350 | 3 115 € | 3,97 | 2,31 |
| American Association for the Advancement of Science (AAAS) | 166 | 3 085 € | 4,97 | 3,36 |
| MyJove Corporation | 110 | 3 081 € | 0,33 | 0,26 |
| The Company of Biologists | 414 | 3 017 € | 1,76 | 1,08 |
| The Endocrine Society | 135 | 3 017 € | 2,09 | 1,42 |
| Society for Neuroscience | 233 | 2 982 € | 2,53 | 1,56 |
| The American Association of Immunologists | 87 | 2 976 € | 1,08 | 1,06 |
| Mary Ann Liebert Inc | 117 | 2 872 € | 1,21 | 1,08 |
| Elsevier BV | 12534 | 2 855 € | 1,99 | 1,99 |

*Correlation test*

We performed a Spearman correlation test on four variables at publisher level: publications number (pub_nbr), APCs average (APCs_avg), MNIJ and MNCS. The results are presented in figure 4.

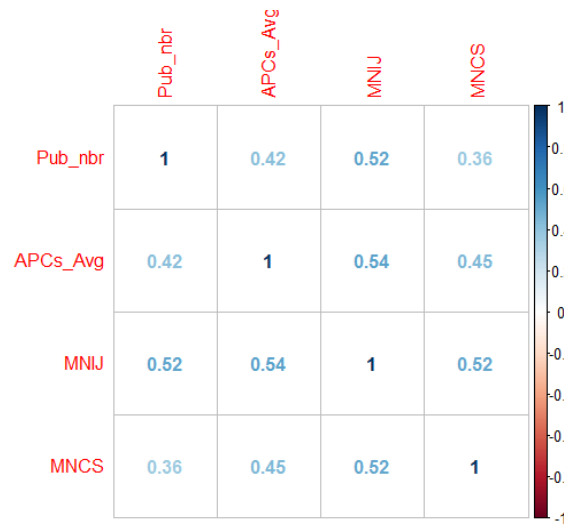

**Figure 4: Spearman correlation matrix**

All coefficients are significant at 5%. We observe that the number of publications is more correlated with the MNIJ (0.52) than with the MNCS (0.36). This means that within the OpenAPC database there are small publishers whose publications have high citations scores and large publishers with low citations scores. We also note that the number of publications per publisher is moderately correlated with the amount of APCs. This would mean that there is no evidence of relationship between the size of publisher and the amount of APCs.

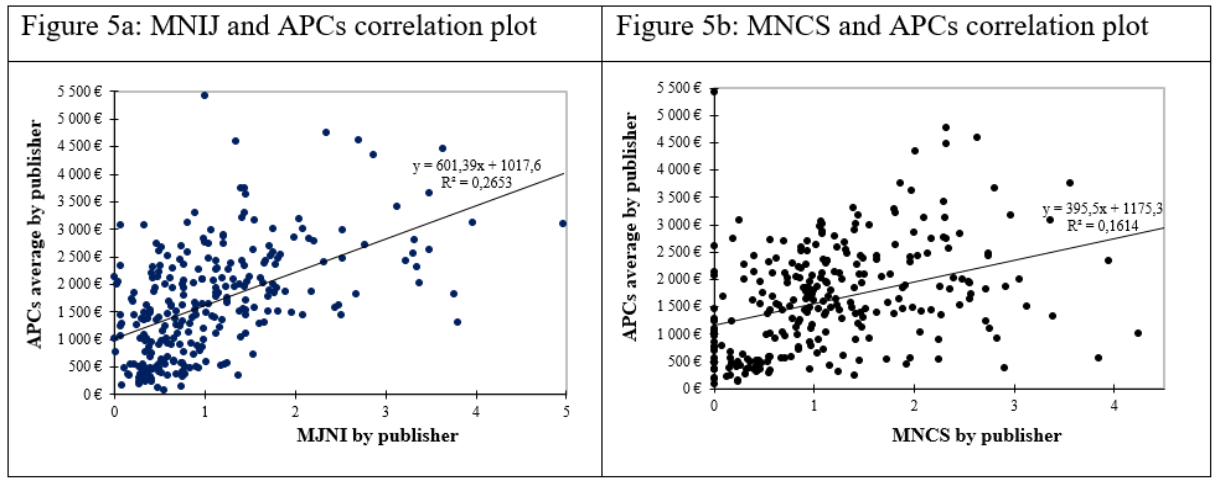

**Figure 5: APCs average per publication**

Figure 3 shows correlation plots between on the one hand MNIJ and APCs (figure 5a), and on the other hand the MNCS and the APCs (figure 5b). The correlation between MNIJ and the amount of APCs is much higher (0.54 against 0.45). This shows that publishers take the quality into account when pricing their journals. However, this prices does not necessarily translate into impact. As long as the APCs are only moderately correlated with the MNCS.

*Regression results*

Table 3 summarizes Tobit regression results for explaining NCS by the amount of APCs. Regression was carried out in two stages. First, only the explanatory variable was integrated (*Log APC* per publication - model 1). Then, control variables were added (model 2).

**Table 3: Tobit maximum likelihood estimation for log NCS**

| Variables_type | Variables | Model_1 | | Model_2 | |
|---|---|---|---|---|---|
| | | Coefficient | Pr(>\|z\|) | Coefficient | Pr(>\|z\|) |
| Explanatory | $Log\ (APCs)$ | 0.22*** | <2.22e$^{-16}$ | 0.012*** | 5.25e$^{-05}$ |
| Control | $Log(journal\_impact)$ | - | - | 0.471*** | <2e$^{-16}$ |
| | $Log(countries\_nbr)$ | - | - | 0.100*** | <2e$^{-16}$ |
| | $Is\_hybrid$ | - | - | 0.204*** | <2e$^{-16}$ |
| Model statistics | Wald-statistic | 2667 | <2.22e$^{-16}$ | 1.238e$^{+04}$ | <2.22e$^{-16}$ |
| | Log-likelihood | -8.063e$^{+04}$ | <2.22e$^{-16}$ | -7.607e$^{+04}$ | <2.22e$^{-16}$ |
| | #publications | 83,753 | | | |
| | #Left-censored | 11907 | | | |
| | #Uncensored | 71846 | | | |
| | #Right-censored | 0 | | | |

*\*\*\*Significant at 1%*

Table 3 shows that when control variables are not taken into account, the amount of APCs strongly impacts citation score (model 1). Once control variables are integrated, the APCs amount impact drops significantly. However, it can be seen that the amount of APCs has a positive impact on citations. Another interesting result is the impact of hybrid journals. Thus, if the journal is hybrid, citations score is higher. In other words, OA articles published in hybrid journals are generally more cited than OA articles published in 100% APCs journals. This is to be expected, given that not all well-established journals in the market have adopted a fully OA business model (Traag and Waltman, 2019). On the other hand, the main large publishers have massively integrated the hybrid model from 2013 (Besancenot and Vranceanu, 2017). In contrast, many fully OA journals are recently created journals that still have not built such a strong reputation for quality (some might even be aiming for average-level reputation and impact if this maximizes income – see (Traag and Waltman, 2019)). Hybrid journals are therefore more likely to be, at moment, in a virtuous circle where they receive higher quality manuscripts than fully OA journals, which translates to higher NCS of the published articles.

**Conclusion and discussion**

Through this article, we seek to analyze relationship between APCs and academic impact. Based on a large sample of 83,752 publications our study empirically verifies the belief that if we pay dearly for publication, impact of publication would necessarily be high. This belief stems from the fact that an author or an institution may think that all publishers who charge a high price for APCs and indexed in international databases like WoS, necessarily have a high academic quality. Our results show that contrary to this belief, paying dearly does not necessarily increase impact of publications. First, large publishers with high impact are not the most expensive in terms of APCs. Second, publishers with highest APCs are not necessarily the bests in terms of impact. Correlation between APCs and impact is moderate.

Otherwise, in the econometric analysis we have shown that publication quality is strongly determined by journal quality in which it is published. This result agrees with several studies which show it empirically (Waltman and Traag, 2017; Maddi, Larivière and Gingras, 2019). International collaboration also plays an important role in citations score. This result is also consistent with literature (Larivière *et al.*, 2015).

Another interesting result relates to the impact of hybrid journals versus 100% APCs journals. The regression results indicate that if the journal is hybrid, the NCS is stronger than if it is fully open. This result is consistent with the study of (Schönfelder, 2020) on the same database (OpenAPC) which showed that journal's impact and hybrid status are the most important factors for the level of APCs.

Our results have several implications for public policy and authors choices when it comes to submit their publications. First, the strong interest for OA had an immediate effect on the publishing market. Prices of OA publications have increased exponentially. This increase is disproportionate to the academic impact. The impact of publications for which authors have dearly paid is no better than that of publications with low APCs. Impact may even be lower. We also showed that some publishers are taking advantage of OA movement to demand high APCs, while their academic impact is very low. Finally, our results suggest that, for authors, APCs should not be used as an indicator for journals selection for submission. For institutions, for efficient management, it is important to be attentive to journals quality before granting funds for OA publication.